\documentclass[tighten,times,twocolumn]{aastex701}

\graphicspath{{./}{figures/}}

\usepackage{amsmath}
\usepackage{mathtools}
\usepackage{upgreek}

\newcommand{\appropto}{\mathrel{\vcenter{
  \offinterlineskip\halign{\hfil$##$\cr
    \propto\cr\noalign{\kern2pt}\sim\cr\noalign{\kern-2pt}}}}}

\usepackage{hyperref}
\hypersetup{ bookmarks = true }
\usepackage{xspace}
\usepackage{soul,xcolor}
\usepackage{amsmath}
\usepackage{multirow}
\usepackage{ulem}
\usepackage{soul}

\usepackage{gensymb}
\usepackage{graphicx}
\usepackage[enable]{easy-todo}
\usepackage[normalem]{ulem}

\usepackage{comment}

\usepackage{savesym}
\savesymbol{tablenum}
\usepackage{siunitx}
\restoresymbol{SIX}{tablenum}
\DeclareSIUnit\degr{deg}
\DeclareSIUnit\parsec{pc}
\DeclareSIUnit\dBm{dBm}
\DeclareSIUnit\jansky{Jy}
\DeclareSIUnit\beam{beam}

\definecolor{ucol}{rgb}{.6,0,0}
\definecolor{webgreen}{rgb}{0,.5,0}
\definecolor{halfgray}{gray}{0.55}

\usepackage[capitalize]{cleveref}

\newcommand{\tcm}{21\,cm\xspace}  

\newcommand{\quotes}[1]{``#1''}

\newcommand{\hi}{\rm H{\textsc{\romannumeral 1}}}
\newcommand{\mgii}{\rm Mg{\textsc{\romannumeral 2}}}

\newcommand{\nhi}{{N_{\rm H{\textsc{\romannumeral 1}}}}}

\defcitealias{2022ApJS..261...29C}{CHIME Collaboration et~al.}
\defcitealias{2018ApJ...863...48C}{CHIME/FRB Collaboration et~al.}
\defcitealias{2021ApJS..255....5C}{CHIME/Pulsar Collaboration et~al.}
\defcitealias{planckcollaborationPlanck2018ResultsVI2020}{Planck Collaboration et~al.}

\shorttitle{CHIME Absorber Detection}

\begin{document}
\setstcolor{red}

\title{Discovery of a \tcm absorption system at z=2.327 with CHIME}
\begin{abstract}
We report the detection of a new \tcm absorption system associated with the radio source NVSS~J164725+375218 at a redshift of $z=2.327$, identified through a pilot survey conducted by the Canadian Hydrogen Intensity Mapping Experiment (CHIME). This is the fifth detection of an associated system at $z > 2$. By analyzing a subset of available data, we conduct a spectrally blind survey for \tcm absorption systems within the redshift range of 0.78 to 2.55 along 202 lines of sight toward known sources in the declination range of $35{\degree}$ to $60{\degree}$. We detect three \tcm absorbers:  two previously known intervening systems and one newly discovered associated system. By fitting the absorption profiles with models containing one to three Gaussian components and selecting the best model using the Bayesian information criterion, we estimate the optical depth, velocity-integrated optical depth, and the ratio between the \hi\ column density and the spin temperature of the absorption systems. 
These results demonstrate CHIME's ability to discover new absorbers, even in a small subset of its full dataset. 
\end{abstract}






\shortauthors{The CHIME Collaboration}

\newcommand{\UBC}{Department of Physics and Astronomy, University of British Columbia, 6224 Agricultural Road, Vancouver, BC V6T 1Z1 Canada}
\newcommand{\CITA}{Canadian Institute for Theoretical Astrophysics, University of Toronto, 60 St. George Street, Toronto, ON M5S 3H8, Canada}
\newcommand{\ASTRON}{ASTRON, The Netherlands Institute for Radio Astronomy, Oude Hoogeveensedijk 4, Dwingeloo, 7991 PD, The Netherlands}

\newcommand{\ASU}{Department of Physics, Arizona State University, Tempe, AZ 85287, USA}
\newcommand{\MITK}{MIT Kavli Institute for Astrophysics and Space Research, Massachusetts Institute of Technology, 77 Massachusetts Ave, Cambridge, MA 02139, USA}
\newcommand{\MITP}{Department of Physics, Massachusetts Institute of Technology, 77 Massachusetts Ave, Cambridge, MA 02139, USA}
\newcommand{\UBCO}{Department of Computer Science, Math, Physics, \& Statistics, University of British Columbia, Okanagan Campus, Kelowna, BC V1V 1V7, Canada}
\newcommand{\DRAO}{Dominion Radio Astrophysical Observatory, Herzberg Research Centre for Astronomy and Astrophysics, National Research Council Canada, PO Box 248, Penticton, BC V2A 6J9, Canada}
\newcommand{\SLAC}{SLAC National Accelerator Laboratory, 2575 Sand Hill Road, Menlo Park, California 94025, USA}
\newcommand{\KIPAC}{Kavli Institute for Particle Astrophysics \& Cosmology, 452 Lomita Mall, Stanford, CA 94305, USA}
\newcommand{\ASIAA}{Institute of Astronomy and Astrophysics, Academia Sinica, 11F of AS/NTU Astronomy-Mathematics Building, No.1, Sec. 4, Roosevelt Rd, Taipei 10617, Taiwan, R.O.C.}
\newcommand{\DI}{Dunlap Institute for Astronomy and Astrophysics, 50 St. George Street, University of Toronto, ON M5S 3H4, Canada}
\newcommand{\DAA}{David A. Dunlap Department of Astronomy and Astrophysics, 50 St. George Street, University of Toronto, ON M5S 3H4, Canada}
\newcommand{\PI}{Perimeter Institute for Theoretical Physics, 31 Caroline Street N, Waterloo, ON N25 2YL, Canada}
\newcommand{\CIFAR}{Canadian Institute for Advanced Research, 180 Dundas St West, Toronto, ON M5G 1Z8, Canada}
\newcommand{\MU}{Department of Physics, McGill University, 3600 rue University, Montr\'eal, QC H3A 2T8, Canada}
\newcommand{\YU}{Department of Physics, Yale University, New Haven, CT 06520, USA}
\newcommand{\TSI}{Trottier Space Institute, McGill University, 3550 rue University, Montr\'eal, QC H3A 2A7, Canada}
\newcommand{\KAI}{Kapteyn Astronomical Institute, University of Groningen, PO Box 800, 9700 AV Groningen, The Netherlands}

\collaboration{100}{The CHIME Collaboration}

\author[orcid=0000-0001-6523-9029, gname=Mandana, sname=Amiri]{Mandana Amiri}
\affiliation{\UBC}
\email{mandana@phas.ubc.ca}

\author[orcid=0000-0002-7758-9859, gname=Arnab, sname=Chakraborty]{Arnab Chakraborty}
\affiliation{\MU}
\email{arnab.chakraborty2@mail.mcgill.ca}

\author[orcid=0000-0002-0190-2271, gname=Simon, sname=Foreman]{Simon Foreman}
\email{simon.foreman@asu.edu}
\affiliation{\ASU}

\author[orcid=0000-0002-1760-0868,gname=Mark,sname=Halpern]{Mark Halpern}
\affiliation{\UBC}
\email{halpern@physics.ubc.ca}

\author[orcid=0000-0001-7301-5666,gname=Alex S.,sname=Hill]{Alex S. Hill}
\email{alex.hill@ubc.ca}
\affiliation{\UBCO}
\affiliation{\DRAO}

\author[orcid=0000-0002-4241-8320,gname=Gary,sname=Hinshaw]{Gary Hinshaw}
\affiliation{\UBC}
\email{hinshaw@phas.ubc.ca}

\author[orcid=0000-0003-4887-8114,gname=Carolin,sname= H\"ofer]{Carolin  H\"ofer}
\affiliation{\UBC}
\affiliation{\KAI}
\email{hofer@astro.rug.nl}

\author[orcid=0000-0003-4179-4073, gname=Albin, sname=Joseph]{Albin Joseph}
\affiliation{\ASU}
\email{ajosep52@asu.edu}

\author[orcid=0000-0001-8064-6116, gname=Joshua, sname=MacEachern]{Joshua MacEachern}
\affiliation{\UBC}
\email{maceachern@phas.ubc.ca}

\author[orcid=0000-0002-4279-6946, gname=Kiyoshi W., sname=Masui]{Kiyoshi W. Masui}
\email{kmasui@mit.edu}
\affiliation{\MITK}
\affiliation{\MITP}

\author[orcid=0000-0002-0772-9326,gname=Juan,sname=Mena-Parra]{Juan Mena-Parra}
\affiliation{\DAA}
\affiliation{\DI}
\email{juan.menaparra@utoronto.ca}
\correspondingauthor{Arash Mirhosseini}
\author[orcid=0000-0002-2626-5985, gname=Arash, sname=Mirhosseini]{Arash Mirhosseini}
\affiliation{\UBC}
\email[show]{arashmirhosseini@phas.ubc.ca}

\author[orcid=0000-0003-2155-9578, gname=Ue-Li, sname=Pen]{Ue-Li Pen}
\affiliation{\DI}
\affiliation{\ASIAA}
\affiliation{\CITA}
\affiliation{\PI}
\email{pen@cita.utoronto.ca}

\author[orcid=0000-0002-9516-3245, gname=Tristan, sname=Pinsonneault-Marotte]{Tristan Pinsonneault-Marotte}
\affiliation{\SLAC}
\affiliation{\KIPAC}
\email{tristpm@stanford.edu}

\author[orcid=0000-0001-6967-7253,gname=Alex, sname=Reda]{Alex Reda}
\affiliation{\YU}
\email{alex.reda@yale.edu}

\author[orcid=0000-0002-4543-4588, gname=J. Richard, sname=Shaw]{J. Richard Shaw}
\affiliation{\UBC}
\email{richard@phas.ubc.ca}

\author[orcid=0000-0003-2631-6217,gname=Seth,sname=Siegel]{Seth R. Siegel}
\affiliation{\MU}
\affiliation{\PI}
\affiliation{\TSI}
\email{ssiegel@perimeterinstitute.ca}

\author[orcid=0009-0003-4114-1301, gname=Yukari, sname=Uchibori]{Yukari Uchibori}
\affiliation{\UBC}
\email{yukariu@phas.ubc.ca}

\author[orcid=0009-0005-1033-3292, gname=Rik, sname=van Lieshout]{Rik van Lieshout}
\email{lieshout@astron.nl}
\affiliation{\CITA}
\affiliation{\ASTRON}

\author[orcid=0000-0002-1491-3738, gname=Haochen, sname=Wang]{Haochen Wang}
\email{hcwang96@mit.edu}
\affiliation{\MITK}
\affiliation{\MITP}

\author[orcid=0000-0001-7314-9496,gname=Dallas, sname=Wulf]{Dallas Wulf}
\affiliation{\MU}
\affiliation{\TSI}
\email{dallas.wulf@mcgill.ca}
%
%
%


\section*{}
~
~
~
~
~
~
\section{Introduction}
\label{sec:intro}
\tcm absorption systems are foreground gas clouds that absorb flux from a background radio source at a wavelength of \tcm, corresponding to the hyperfine transition of neutral atomic hydrogen (\hi). The detectability of \tcm absorption depends on the \hi\ column density and spin temperature of the foreground gas, along with the brightness of the background radio source. This makes it a complement to \hi\ emission observations at higher redshifts, where direct detection of emission is difficult because the strength of the emission line decreases with the square of the distance from the emitting gas. The \tcm absorption line serves as a powerful probe of cold gas in the interstellar medium (ISM), providing insights into the structure and evolution of galaxies across cosmic time. In addition to studying galaxies, \tcm absorbers can be utilized to investigate the spin temperature of \hi\ gas \citep{kanekarSpinTemperatureHighredshift2014,allisonStatisticalMeasurementSpin2021}, explore the temporal and spatial variations of fundamental constants \citep{wolfeLimitsVariationFundamental1976, carilliAstronomicalConstraintsCosmic2000, chengalurConstrainingVariationFundamental2003} and directly measure cosmic acceleration \citep{sandageChangeRedshiftApparent1962,darlingDIRECTMEASUREMENTCOSMIC2012,yuMethodDirectMeasurement2014}.

\tcm absorbers are classified into two main types: associated (or intrinsic) absorbers, where the absorbing gas is gravitationally bound
to the background radio source, and intervening absorbers, which arise from gas clouds in the disks or halos of foreground galaxies along the line of sight to a background radio source. Associated absorbers are traditionally defined to be systems within 3000 $\unit{\km}$ $\unit{\second^{-1}}$ of the emission line redshift of their host source \citep{ellisonCORALSSurveyII2002}. Since associated absorbers probe the neutral gas associated with active galactic nuclei (AGNs), they can help constrain models of AGN formation and evolution. Conversely, intervening absorbers can be used to study the properties of the ISM in high-redshift galaxies \citep{morgantiInterstellarCircumnuclearMedium2018, duttaColdNeutralHydrogen2019}. 

\par The current sample of \tcm absorbers, primarily identified from catalogs of damped Lyman-alpha absorbers (DLAs) and \mgii\ absorbers, is limited by observational biases. Optical surveys for DLAs, which rely on the Lyman-alpha transition in the ultraviolet, are restricted to redshifts $ z \gtrsim 1.7 $ due to atmospheric cut-off, and expensive space-based observations are needed to probe these systems at lower redshifts. Additionally, dust in high-column-density, metal-rich DLAs can obscure background quasars, potentially biasing optical samples against such systems and contributing to the observed lack of metallicity evolution in DLAs.

\par An unbiased blind radio survey for \tcm absorption systems overcomes these limitations, as radio waves are unaffected by dust and atmospheric absorption, allowing for a uniform census of cold hydrogen gas across a wide redshift range. Thanks to the enhanced bandwidth coverage and sensitivity of radio telescopes, several blind surveys for \tcm absorbers have been conducted in recent years, including the Five-hundred-meter Aperture Spherical Radio Telescope All Sky \hi\ survey (FASHI), covering declinations between $-16^{\degree}$ to $ 66^{\degree}$ at $z<0.42 $ \citep{zhangFASHIUntargetedSurvey2025}; the First Large Absorption Survey in \hi\ (FLASH), which covers the sky south of declination $ 40^{\degree} $ in the redshift range $0.42<z< 1.0$ \citep{allisonFLASHEarlyScience2020,allisonFirstLargeAbsorption2022a}; and the MeerKAT Absorption Line Survey (MALS), designed to conduct an unbiased survey of \hi\ and OH absorption at redshift $ z < 2 $ in the declination range of $-40\degree<\delta<30\degree$ \citep{guptaMeerKATAbsorptionLine2018}. 

\par To date, only ten detections of \tcm absorbers at $z>2$ have been reported, and four of these are associated absorbers \citep{guptaEvolutionColdGas2021,adityaUGMRTDetectionAssociated2020,mooreNeutralHydrogen211998,usonRadioDetectionsNeutral1991}. 
\cite{curran21CmABSORPTION2010} propose that the scarcity of associated \tcm absorbers may result from the bias in optically selected surveys, which often target luminous AGNs at high redshifts. These AGNs can fully ionize the surrounding neutral hydrogen, leading to the non-detection of the \tcm absorption line. 
\par The Canadian Hydrogen Intensity Mapping Experiment (CHIME) Absorber project aims to increase the sample size of \tcm absorbers at high redshifts by conducting a wide-area blind survey in the redshift range $0.78 < z < 2.55$. The catalog of \tcm absorbers in this redshift range can then be utilized to characterize the cold gas in galaxies and the environment of AGNs. In this paper, we present the results of a pilot spectrally blind search for \tcm absorbers using the data collected by the CHIME Absorber backend, which involved analyzing four datasets, each covering one month of data. This search resulted in the detection of a new absorber associated with the radio source NVSS~J164725+375218 at redshift 2.327, as well as two previously known intervening absorbers. 

\section{Methods}
\subsection{Instrument}

CHIME (\citetalias{2022ApJS..261...29C} \citeyear{2022ApJS..261...29C}) is a transit radio telescope located at the Dominion Radio Astrophysical Observatory near Penticton, BC, Canada. It consists of four cylindrical parabolic reflectors of $20\,\unit{\meter} \times 100\,\unit{\meter}$ each, aligned along the north-south direction. Each reflector contains a linear array of 256 dual-polarization cloverleaf feeds \citep{2017arXiv170808521D} connected to low-noise amplifiers \citep{Davis_2012},  band-pass filters, and analog-to-digital converters. The telescope is sensitive to frequencies from 400\,\unit{\mega\hertz} to 800\,\unit{\mega\hertz}, which corresponds to the \hi\ line at redshifts between 2.55 and 0.78.  

\par A time-domain data stream from each antenna arrives in CHIME's F-engine, where it is channelized in field-programmable gate arrays by performing a polyphase filter bank (PFB, \citealt{polyphase-filter-bank}) using a fast Fourier transform (FFT) on 2048 samples at a time, giving frames of 2.56\,\unit{\micro\second} with a spectral resolution of 391\,\unit{\kilo\hertz}. After applying a scaling factor and phase offset for each frequency channel for optimal compression, these are rounded to 4+4\,bit complex values for transmission to CHIME's GPU-based X-engine \citep{2018JAI.....750008M}. 

\par In the GPU nodes, the data stream for CHIME's Absorber and FRB backends (\citetalias{2018ApJ...863...48C} \citeyear{2018ApJ...863...48C}) splits off from those for the Intensity Mapping (\citetalias{2022ApJS..261...29C} \citeyear{2022ApJS..261...29C}) and Pulsar (\citetalias{2021ApJS..255....5C} \citeyear{2021ApJS..255....5C}) backends, which have different needs. We implement a hybrid beamforming pipeline that generates 1024 beams organized as a $256\times4$ grid, with 256 FFT-formed beams in the North-South (NS) direction and 4 beams formed via exact phasing in the East-West (EW) direction \citep{2017ursi.confE...4N}.  In the NS direction, the beam pointing covers declinations north of $-10\degree$. In the EW direction, the beams point $(-0.4\degree, 0.0\degree, 0.4\degree, 0.8\degree)$ away from the meridian.

\par \tcm absorption features are usually narrower than CHIME's original spectral resolution, so data for the absorber search need a higher resolution than CHIME's native 391\,kHz channels. For each beam and frequency channel, 128 successive frames of 2.56\,\unit{\micro\second} are accumulated and Fourier-transformed to achieve a spectral resolution of 3\,\unit{\kilo \hertz}, corresponding to a velocity resolution between 1.14 to 2.29 $\unit{\km.\s^{-1}}$ across the CHIME frequency band. These data are squared, integrated to 10\,\unit{\second}, and written to disk for archiving and analysis. 

The ability of the CHIME Absorber system to continuously scan a large part of the sky at a sufficiently high spectral resolution makes it an excellent instrument for discovering new \tcm absorbers.

\subsection{Observations and data processing}
\label{method}
The CHIME Absorber backend began collecting data in February 2021, and we currently have approximately four years of data. In this pilot program, we process a subset of these data, allowing us to understand the characteristics of the pipeline output before conducting a more comprehensive search for \tcm absorption lines by processing additional data.

\par The objective of this pilot survey is to identify \tcm absorbers within a redshift range of 0.78 to 2.55 toward radio sources brighter than 400 mJy at 1.4\,\unit{\giga\hertz} with redshift $z > 0.78$, in the declination range of $35{\degree}$ to $60{\degree}$.  We processed four months of data, covering periods from mid-March to mid-April 2021 and 2023 (hereafter referred to as the April datasets), and December 2021 and 2023. During these 4 months, the instrument was more stable than the other periods in our archive with little downtime and fewer transient radio frequency interference (RFI) events. The April and December datasets are eight months apart, enabling us to reject many false positives using the expected barycentric shift between the April and December data caused by Earth’s orbital motion. 
\par Each dataset is processed individually using the CHIME Absorber pipeline, which includes RFI excision, sidereal re-gridding (which transforms the time-domain data onto a fixed grid in local Earth rotation angle ranging from $0{\degree}$ to $360{\degree}$), and correcting for the PFB passband shape which becomes evident as the data are re-transformed to 3\,\unit{\kilo \hertz}.

The band-shape correction is performed in two steps: first, each coarse channel is divided by a PFB model, which is computed from the square of the Fourier transform of the windowing function used in the PFB for channelizing the data. Second, we subtract the spectrum of each pixel in the sky map from the average of its neighboring pixels in the right ascension direction. This helps remove truncation noise bias caused by 4+4\,bit truncation in the F-engine, which is not accounted for in our PFB model. This subtraction procedure works under the assumption that the bias remains constant over the time samples corresponding to neighboring pixels in right ascension, which is valid when no gain updates occur between them. Finally, maps made on each sidereal day in a month are averaged together. The output is a 3D map of the sky for each dataset, comprising spectra from over two hundred thousand pixels, which were subsequently saved for analysis.  The integration time for each line of sight is approximately $60\,\unit{\s}$ per day, resulting in a total integration time of about 30 minutes per month.

\par Non-negligible spectral structures that vary between coarse channels remain even after the two-step band-shape correction. Within each coarse channel, these residuals are well-described by a sine function with a wavelength equal to the coarse channel width. To mitigate this, we fit each coarse channel with one full oscillation of a sine function for each line of sight to extract the phase and amplitude. We assume that the phase and amplitude vary slowly from channel to channel and smooth the fitted parameters using the arPLS method \citep{baekBaselineCorrectionUsing2015}. We then subtract the resulting sine function, with the smoothed amplitude and phase, from the coarse channel. This approach ensures that outlier features, which include absorption features, remain unaffected.

\par We calculate the average frequency shift caused by the Earth's motion over one month along each line of sight and apply this correction to the monthly spectra to obtain barycentric corrected maps.  
Performing this barycentric correction at a monthly cadence results in loss of signal amplitude, particularly for narrow absorption features at higher frequencies. However, the amplitude loss is minimal for the three detections reported in this work: two are located in the lower portion of CHIME's frequency band and exhibit moderate widths, while the third, though at a higher frequency, is sufficiently broad to remain largely unaffected. We plan to incorporate the barycentric correction task into the daily processing pipeline in our future full search.

\par Our long-term goal is to process all lines of sight to conduct an untargeted and unbiased blind radio survey for \tcm absorbers. 
For this pilot search, we focus on a limited subset of data and select lines of sight near the zenith and with sufficiently bright background sources.  We used the NRAO VLA Sky Survey (NVSS) catalog \citep{condonNRAOVLASky1998} and available redshift information taken from the NASA/IPAC Extragalactic Database (NED)\footnote{\url{https://ned.ipac.caltech.edu/}} to select 202 lines of sight from the processed output. These lines of sight were chosen based on having a background source brighter than 400\,mJy at 1.4\,\unit{\giga\hertz} and a redshift greater than 0.78 so that they can act as a background source for the absorbers. Requiring an optical redshift measurement introduces a bias toward the radio sources that are less dusty and exhibit strong UV brightness. Note that unlike targeted \tcm absorption surveys toward DLAs or \mgii\ absorbers, where the redshift of the absorption feature is known from optical surveys and used to guide \hi\ absorption searches in the radio spectrum, our pilot survey represents a spectrally blind search, i.e., we do not use prior knowledge about the \hi\ absorption frequencies of absorbers during the search. 

\subsection{Search for absorption features}
\label{sec:methods:search}

Our search strategy proceeds in three stages: 1)~applying a matched filter to the spectrum of each line of sight and using a likelihood ratio test to identify candidate absorption signals; 2)~rejecting candidates from stage 1 whose line centers in different datasets are inconsistent with the expected seasonal Doppler shift; and 3)~visually inspecting the remaining candidates and rejecting those displaying characteristic features of RFI or other systematics. We describe the details of each stage below.

\par We design a matched filter based on a likelihood ratio test comparing the null hypothesis that the observed data are due to Gaussian noise against the hypothesis that the signal is a genuine absorption-like feature. The logarithm of likelihood ratio values ($\ln \Lambda$) is evaluated by convolving spectra with models consisting of a set of Gaussian templates with full width at half maximum (FWHM) ranging from 35\,$\unit{\kilo\hertz}$ to 175\,$\unit{\kilo\hertz}$. We identify negative-amplitude features with $\ln \Lambda>20$ as absorber candidates. The threshold of $\ln \Lambda = 20$ was determined empirically through iterative testing to balance sensitivity to genuine absorption features against the computational and human effort required for visual inspection of false positives. Since residual band-shapes are stronger toward bright background sources (which comprise our target sample of sources above 400\,mJy at 1.4\,\unit{\giga\hertz}), a lower threshold produces too many false positives for practical visual inspection. Conversely, a higher threshold risks excluding genuine absorption features. We therefore chose this threshold to yield a manageable number of candidates for thorough visual inspection while maintaining sensitivity to real systems.

\par There are many features in the spectra, such as spectral jumps, RFI, and residual band-shape, that are not accounted for in our null model. These features generate numerous candidates we identify as false positives, with the majority originating from systematic effects (primarily residual band-shape)  that persist across frequency and time, and unlike astrophysical signals, they don't exhibit a Doppler shift across datasets. To reduce the number of false positives, we implement a consistency check, requiring that the peak of an absorption feature agrees to within 18\,\unit{\kilo\hertz} in the barycentric reference frame across all available datasets. This criterion successfully eliminates the majority of candidates with the properties described above. Nevertheless, spectral features that persist across all datasets can still pass the validation criteria and generate false positives. Such false positives occur predominantly in the sight lines with large residual band shape amplitudes and at frequencies contaminated by broadband RFI.

\par Following the validation described above, we visually inspect the remaining candidate spectra. We discard any features with morphologies consistent with RFI, large residual band-shape errors, or other instrumental systematics. We also discard features that are not consistently present in all EW beams. In Appendix~\ref{sec:falsepositives}, we provide an additional statistical test of the candidates that undergo visual inspection, which verifies that our inspection procedure successfully identifies statistical outliers that can be interpreted as genuine absorption systems.

\subsection{Continuum correction}
\label{sec:methods:continuum}

\par The large angular size of CHIME's synthesized beams (between $0.25\degree$ and $0.5\degree$ depending on frequency) presents a challenge for accurate continuum estimation, as multiple radio sources might fall within a single beam and contribute to the continuum level. To account for this, we implement a continuum correction method that removes the contribution from additional bright sources within the synthesized beam.
\par For each detection, we identify all NVSS sources within the beam and determine their angular positions relative to the beam center. Using the known beam sensitivity pattern, we calculate the contribution of each source to the continuum level by estimating its flux density at the absorption frequency and accounting for the beam response at its location.   We estimate the flux density of each source at the absorption frequency by interpolating from the cataloged 1.4\,\unit{\giga\hertz} and 408\,\unit{\mega\hertz} flux densities assuming a power-law model.  The flux density measurements at 408\,\unit{\mega\hertz} and 1.4\,\unit{\giga\hertz} are derived from the Bologna Sky Survey \citep{ficarraNewBolognaSky1985} and the NVSS catalog \citep{condonNRAOVLASky1998}, respectively.  For sources without a reported flux at 408 MHz, we adopt a conservative spectral index of $-2$ (corresponding to an ultra-steep spectrum), which provides an upper limit on their contribution to the continuum. We note that this power-law assumption may not be accurate for gigahertz-peaked spectrum (GPS) sources, where synchrotron self-absorption causes the spectrum to turn over, typically at frequencies between 0.4 and 5\,\unit{\giga\hertz} \citep{1983A&A...123..107G,2021A&ARv..29....3O}. However, GPS sources with turnover frequencies near 1.4\,\unit{\giga\hertz} represent only $7.4\%$ of the population of compact radio sources \citep{2024A&A...689A.264B}. Moreover, the contaminating sources to which we apply this correction are typically faint and located away from the beam center where the beam sensitivity is significantly reduced, making GPS-related systematics unlikely to significantly affect our continuum estimates. To isolate the contribution from the source producing the absorption feature, we subtract the sensitivity-weighted flux from contaminating sources until the cumulative contribution from remaining sources is below 10\% of the total continuum level. This corrected continuum value is then used in the optical depth calculation described in Section~\ref{properties}. We note that the remaining sources contribute to the confusion noise in our measurements.

\subsection{Physical and observable properties}
\label{properties}
\par Two main observables in \hi\ absorption studies are the fractional change in continuum level due to absorption ($\Delta S/S_0$) and the width of the absorption feature. These observables are used to estimate the observed optical depth and the ratio between \hi\ column density $\nhi$ and the spin temperature $T_\mathrm{s}$ of the absorbing cloud. The optical depth is defined as $\tau = -\ln(1 - \Delta S / (f S_0))$, where $S_0$ and $f$ represent the continuum level and the covering factor of the background source, respectively. In the optically thin regime ($\Delta S / S_0 \lesssim 0.3$), the optical depth can be approximated by $\tau \approx \Delta S / (f S_0)$. The relationship between the true and observed optical depth is $\tau=\tau_{\rm obs}/f$. The \hi\ column density can then be determined by

\begin{equation}
    \nhi = A \times \frac{T_\mathrm{s}}{f} \int \tau_{\rm obs} \, dv,
\end{equation}
where $A=1.823 \times 10^{18}\,\unit{\cm^{-2}\,\kelvin^{-1}\,\km^{-1}\,\second}$ is a constant factor, and $\int \tau_{\rm obs} \, dv$ is the velocity-integrated optical depth \citep{wolfePhysicalConditionsLocations1975}.
\par To obtain numerical estimates of the optical depth and the line width of the absorption systems, we model the absorption spectra as a superposition of Gaussian components. A simple way to interpret multiple Gaussian components in an absorption spectrum is to treat them as being individual absorbing clouds along the line of sight. We fit the spectra with models having one to three Gaussian components. Then, the best model is selected using the Bayesian Information Criterion (BIC; \citealt{schwarzEstimatingDimensionModel1978}). The BIC is defined as
\begin{equation}
    \text{BIC}=k \ln(n) - 2\ln(\hat{L})
\end{equation}
where  $k$ is the number of parameters in the model, $n$ is the number of data points, and 
$\hat{L}$ is the maximum likelihood value of the model. Models with a lower BIC value are preferred. The fits give the peak absorption frequency $\nu_{\rm abs}$, peak observed optical depth $\tau_{\rm peak}$, and FWHM of each component, which are then used to estimate the quantities $\int \tau_{\rm obs} \, dv$ and $f\nhi/T_\mathrm{s}$ for each individual component, as well as for the entire absorption feature. For absorbers with multiple Gaussian components, the total absorption frequency is estimated by taking the weighted average, using the velocity-integrated optical depths as weights.

\section{results and discussion}

\label{sec:results}

Our pilot spectrally blind search has yielded three detections of \tcm absorption systems: two known intervening systems and one new associated absorber. The peak absorption frequency $\nu_{\rm abs}$, the peak observed optical depth $\tau_{\rm peak}$, and FWHM of each Gaussian component, and of the full line, are listed in Table~\ref{summary_table}.

\begin{table*}[t!]
\caption{Characteristics of three \tcm absorbers detected with CHIME.}
\hspace{-25mm}
\vspace{1mm}
\begin{tabular}{cccccccccc}
\hline
\hline

 NVSS name & Type & Comp.&$\nu_{\rm abs}$ & $z$ &FWHM& FWHM&$\tau_{\rm peak}$ &$\int{\tau dv}$  & $\frac{f\nhi}{T_\mathrm{s}}$ \\ 
 & & & (\unit{\mega\hertz}) & &(\unit{\kilo\hertz}) & (\unit{\km.\s^{-1}}) & & (\unit{\km.\s^{-1}}) & (\unit{\cm^{-2}.\kelvin^{-1}}) \\
 \hline
  \multirow{3}{*}{NVSS~J164725+375218} & \multirow{3}{*}{A}&1& 426.859(1)& 2.32758(1)& 58(4) &41(3) &0.067(2)&2.9(2) &5.4(4)$\times 10^{18}$\\
  &&2&426.915(1) &2.32714(1)&37(3)&26(2)&0.054(3)&1.5(1) &2.7(3)$\times 10^{18}$ \\  
  &&Total& 426.878(1)  & 2.32743(1)  &98(2) &69(1)&0.067(2) & 4.4(2) &8.1(4)$\times 10^{18}$ \\ 
  \hline
  NVSS~J031443+431405 & I&1& 431.764(1)  & 2.28978(1) & 67(3)&46(2)&0.0159(6)&0.78(5) & 1.42(8)$\times 10^{18}$ \\
  \hline
   \multirow{3}{*}{NVSS~J235421+455304} &  \multirow{3}{*}{I}&1& 798.151(18)& 0.77962(4) &112(25)&42(9) &0.16(3)& 5.7(1.2)& {1.0(2)}$\times 10^{19}$\\ 
   
   &&2& 798.238(7)  & 0.77942(2)&90(9)&33(3) &0.31(6) & 13.8(4.1)& 2.6(7)$\times 10^{19}$\\ 

   &&Total& 798.204(11) & {0.77950(2)} &{134(12)}&51{(5)}&0.36(2) & 19.5(3.6)& 3.6(7)$\times 10^{19}$\\ 
   
  \hline
\end{tabular}
\textbf{Notes:} The second column is the absorber type, with \quotes{A} and \quotes{I} standing for Associated and Intervening absorbers, respectively. The third column is numbered Gaussian component. Columns 4, 6, 7 and 8 (frequency $\nu_{\rm abs}$, FWHM in \unit{\kilo\hertz} and \unit{\km.\s^{-1}}, and peak optical depth $\tau_{\rm peak}$) are parameters obtained directly from Gaussian fits to the absorption profiles. The remaining columns (redshift $z$, velocity-integrated optical depth $\int{\tau dv}$, and the $\frac{f\nhi}{T_\mathrm{s}}$ ratio) are quantities calculated from the fitted parameters. Numbers in parentheses show uncertainties in the final digits of the quantities. 
\label{summary_table}
\end{table*}
\subsection{A new associated absorber at $z={}$2.327}
NVSS~J164725+375218 is a compact radio source \citep{machalskiCompactRadioSources1990} whose optical counterpart is identified as a quasar at a redshift of $2.331$ in the SDSS DR16 Quasar Catalog \citep{lykeSloanDigitalSky2020}. This source has a flux density of 608\,mJy at 1.4\,$\unit{\giga\hertz}$ and 1.23\,Jy at 408\,$\unit{\mega\hertz}$, corresponding to a spectral index of $-0.57$, which classifies it as a compact steep spectrum (CSS) radio source. CSS sources are young radio galaxies in an early evolutionary stage, typically hosting dense circumnuclear gas that can give rise to associated \tcm absorption systems.
\par
\begin{figure}[t!]
    \centering
    \includegraphics[width=1\linewidth]{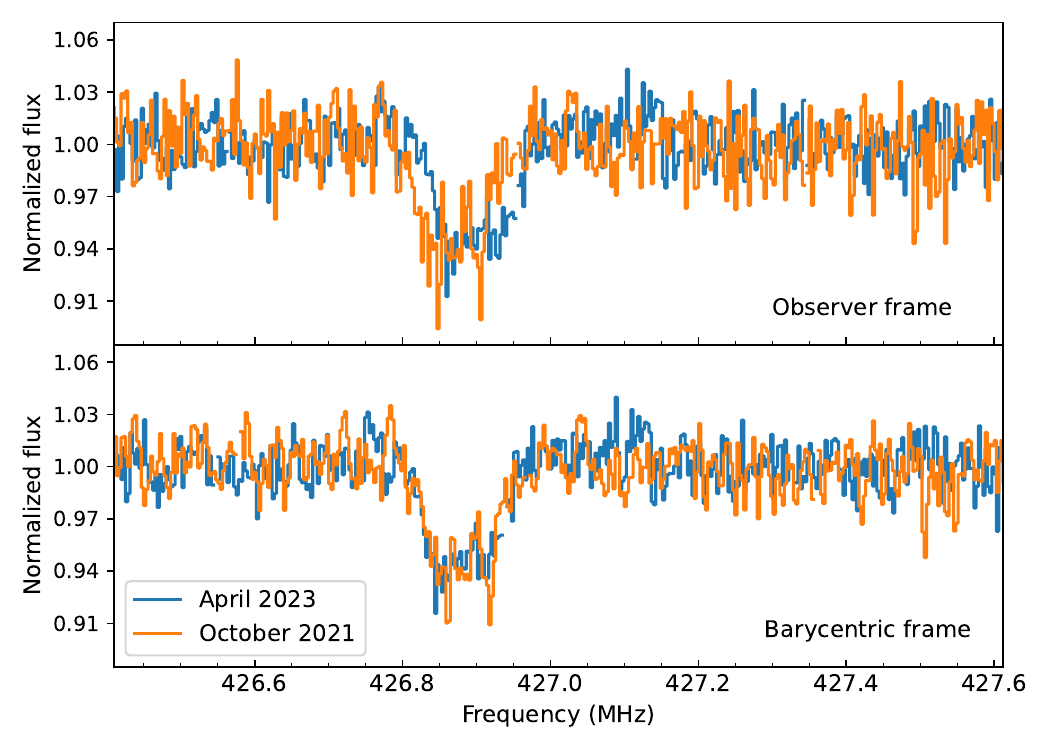}
    \caption{Detection of a \tcm absorption feature in the spectrum of the radio source NVSS~J164725+375218 from October 2021 (orange) and April 2023 (blue) data. Frequencies in the top and bottom panels are presented in the observer and barycentric frames, respectively. The raw spectrum is normalized by the continuum value at the absorption frequency, with the continuum derived by fitting a polynomial of degree 6, excluding 400 \unit{\kilo\hertz} around the absorption feature. Note that the barycentric correction involves interpolating data at the barycentric frequencies, which explains why the shapes of the curves in the two panels are not identical.}
    \label{new-system}
\end{figure}

\begin{figure}[htbp!]
    
    \includegraphics[width=1\linewidth]{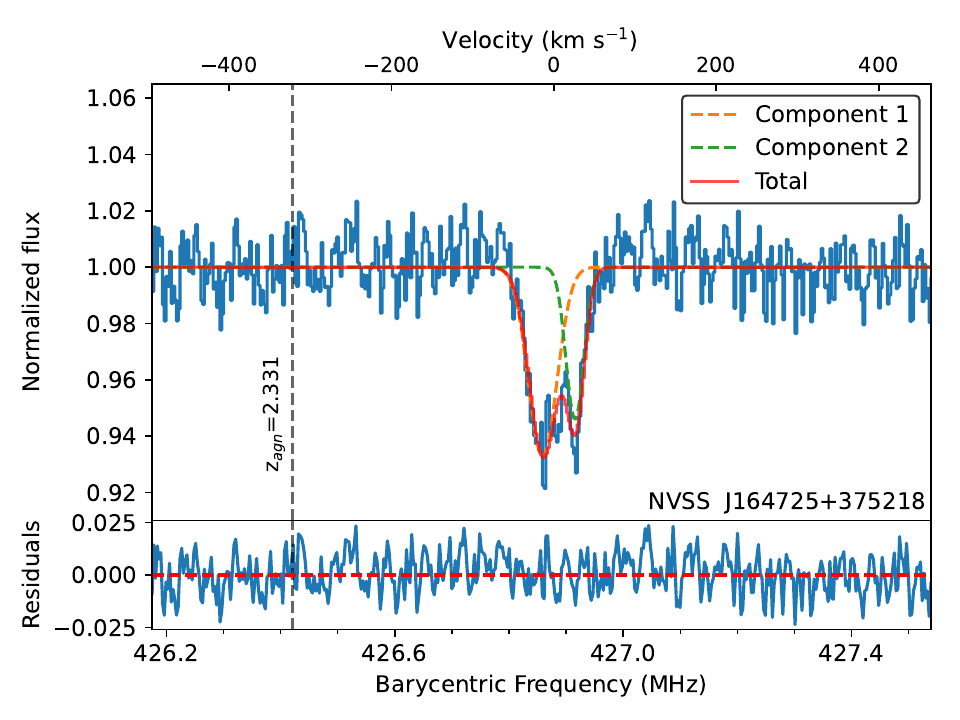}
    \caption{Detection of \hi\ absorption at $z=2.32743$ after combining October 2021 and April 2023 data. The top axis shows the velocity of the absorbing gas relative to the redshift of the absorber (total redshift in Table \ref{summary_table}). The vertical black dotted line corresponds to the redshift of the host radio source. The vertical axis represents the normalized spectrum, which is the ratio of the spectrum to the estimated continuum level at the absorption line's peak.}
    \label{associated-absorber}
\end{figure}

 An absorption-like feature at 426.878\,$\unit{\mega\hertz}$ was identified in the line of sight toward NVSS~J164725+375218 by our search algorithm in both the April 2021 and 2023 datasets. However, this source is masked in the December datasets due to its proximity to the Sun. To test the validity of this candidate, we processed an additional month of data from October 2021, where the same absorption feature is visible at a frequency shifted by 28\,\unit{\kilo\hertz}, consistent with the barycentric shift caused by the Earth's motion toward this source. Figure~\ref{new-system} shows the absorption feature in the October 2021 and April 2023 data in both the observer and barycentric frames. When we apply a barycentric correction, the absorption frequencies of the two spectra are aligned, as expected for an astrophysical source but inconsistent with most sources of RFI or systematic effects. Moreover, this absorption feature shows the correct Doppler shift in all East-West beams in the April and October datasets, leading us to conclude that it is a genuine absorption feature, highly unlikely to be caused by RFI or systematic effects. Figure~\ref{associated-absorber} presents the final barycentric-corrected spectrum after combining the October 2021 and April 2023 data. While this absorption feature is also visible in the April 2021 data, we did not include those data in Figure~\ref{associated-absorber} due to systematic artifacts resulting from band-shape flattening and the presence of RFI in nearby channels.

\par There are another two bright NVSS sources within the same beam as NVSS~J164725+375218. The brightest of these is NVSS~J164800+374429, located 10 arcminutes away from the NVSS~J164725+37521 with a flux density of 700\,mJy at 1.4\,$\unit{\giga\hertz}$ and 2.09\,Jy at 408\,$\unit{\mega\hertz}$, contributing significantly to the continuum value. The other source, NVSS~J164806+380105, has a flux density of 243\,mJy at 1.4\,$\unit{\giga\hertz}$ and 740\,mJy at 408\,$\unit{\mega\hertz}$. This source is 12 arcminutes away from the absorbing source, and while it is fainter than the other contaminating source, it contributes more strongly as it is closer to the beam centre.  To accurately estimate the optical depth, we apply the continuum correction described in Section~\ref{sec:methods:continuum}, accounting for the contribution of NVSS~J164800+374429 and NVSS~J164806+380105 in the beam. Using the spectral indices of NVSS~J164725+375218 and the contaminating sources, we interpolate their flux densities to the absorption frequency and subtract the sensitivity-weighted contributions of the contaminating sources from the total continuum. The resulting optical depth values listed in Table~\ref{summary_table} reflect this correction.
\par
\begin{figure}
    \centering
    \includegraphics[width=1\linewidth]{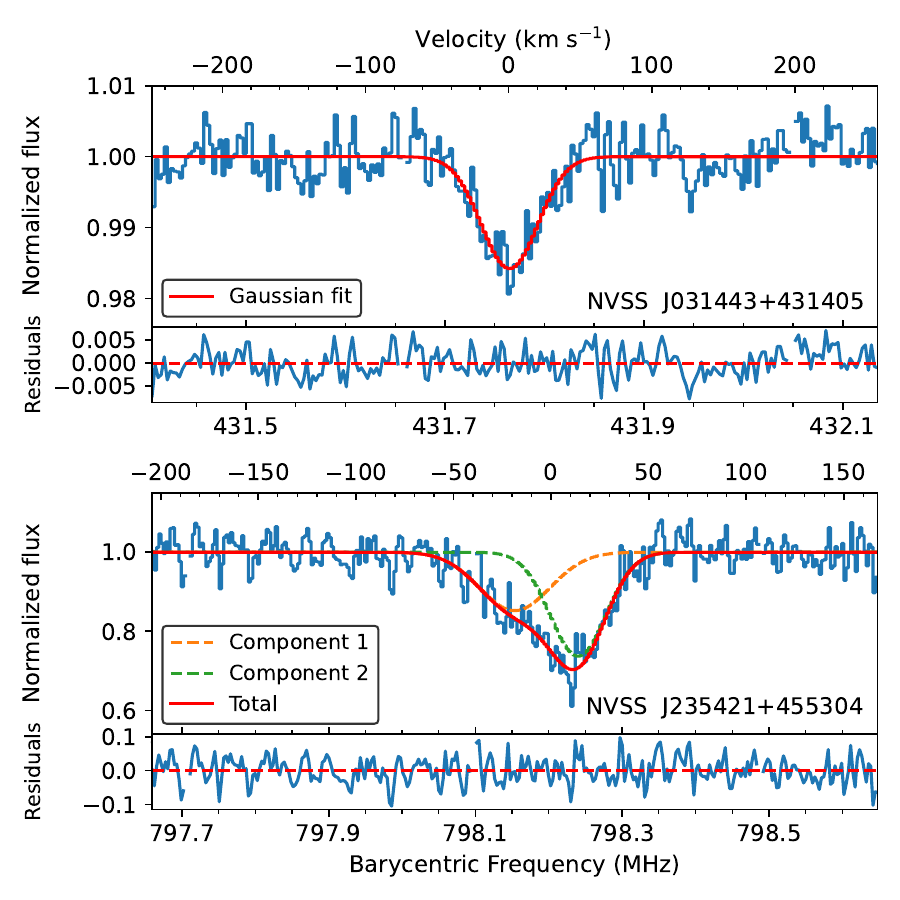}
    \caption{Detection of two known \tcm absorbers. Both spectra utilize December 2021 and 2023 data. The top and bottom axes in each panel are velocity in the rest frame of the absorption system and frequency in the barycentric frame, respectively.}
    \label{intervening-absorbers}
\end{figure}

 The absorption line appears at a frequency of 426.878\,MHz, corresponding to a redshift of 2.32743. Since the relative difference between the redshift of NVSS~J164725+375218 and two components of the absorber is $\sim$\,300\,\unit{\km.\s^{-1}} this absorber is most likely to be associated with this AGN.  This marks the fifth detection of an associated \tcm absorber at a redshift of $z>2$ and the fourth highest redshift of an associated absorber detected to date. 
The absorber is well described by a two-component Gaussian model with a total velocity width of 71\,\unit{\km.\s^{-1}} as shown in Figure~\ref{intervening-absorbers}. Previous studies of associated absorbers, such as \cite{gerebHIAbsorptionZoo2015} and \cite{maccagniKinematicsPhysicalConditions2017} suggest that narrow absorption profiles with width $\lesssim 200\,\unit{\km.\s^{-1}}$ are typically produced by \hi\ clouds rotating in circumnuclear disks, while broader profiles with blueshifted wings are due to gas-rich mergers or strong outflows. The total line width of this absorber, and the low velocity offset compared to the AGN suggest that the absorption lines are likely to be produced by \hi\ clouds rotating in a circumnuclear disk.

\par \citet{curran21CmABSORPTION2010} established an ultraviolet luminosity threshold of approximately $10^{23}\,\unit{\watt\,\hertz^{-1}}$, above which associated \tcm absorption is typically not detected, presumably because the high UV flux ionizes the neutral hydrogen. We estimated the ultraviolet luminosity of the host AGN using SDSS \textit{u}, \textit{g}, \textit{r}, \textit{i}, and \textit{z} magnitudes \citep{lykeSloanDigitalSky2020}. After correction for Galactic extinction, these magnitudes are converted to flux density $F_{\nu}$, and further to a luminosity via $L_{\nu}=4\pi F_{\nu} D_L^2 /(1+z)$, where $D_L$ is the luminosity distance estimated using the $\Lambda$CDM cosmology with parameters from \citetalias{planckcollaborationPlanck2018ResultsVI2020} \citeyearpar{planckcollaborationPlanck2018ResultsVI2020}, and $z$ is the redshift of the radio source. A power law is then fitted to the UV luminosities and the luminosity at the Lyman-$\alpha$ wavelength is estimated to be $L_{121.567} \approx 2 \times 10^{22}\,\unit{\watt\,\hertz^{-1}}$, which is below the threshold, consistent with our detection of associated \hi\ absorption in this source.

\subsection{Detection of two previously known absorbers}

There are a total of 13 known \tcm absorbers in the redshift range covered by CHIME and within the declination range of $35{\degree}$ to $60{\degree}$. Of these 13 absorbers, 5 fall within RFI bands, and 6 are too weak to be detected in this pilot search, with absorption amplitudes ranging from 3\,mJy to 37\,mJy. These amplitudes are below the noise level in one month of our data, which ranges from 20 to 45\,mJy across 400$-$790\,\unit{\mega\hertz}, and increases to about 100\,mJy at frequencies above 790\,\unit{\mega\hertz}. Therefore, we need to process more data to re-detect these weaker absorbers.

However, we have successfully re-detected the remaining two previously known absorbers across all our datasets. The first is an absorption feature at a frequency of 431.765\,$\unit{\mega\hertz}$ toward the source NVSS~J031443+431405, first reported by \cite{yorkDiscovery21cmAbsorption2007}. The absorber itself is at a redshift of 2.29, while the background source for this absorber is a quasar at a redshift of $z \approx 2.87$, with a flux density of 1.3\,Jy at $1.4\,\unit{\giga\hertz}$ and 4.94\,Jy at $408\,\unit{\mega\hertz}$.   Importantly, there are no other NVSS sources within the beam with brightness comparable to NVSS~J031443+431405, allowing for a straightforward continuum estimation. Our derived parameters for this absorber are consistent with the values reported in \cite{yorkDiscovery21cmAbsorption2007}, within the estimated uncertainties. \cite{yorkDiscovery21cmAbsorption2007} conducted optical follow-up observations of this absorber using the Gemini Multi-Object Spectrograph (GMOS) on the Gemini-North telescope, confirming the presence of damped Lyman-alpha absorption at the absorber redshift of 2.29. Their analysis revealed a notably low spin temperature of 138\,\unit{\kelvin}, among the first such measurements in a high-redshift DLA. They concluded that the combination of this low spin temperature, high metallicity, and large velocity spread observed in both the \tcm and metal absorption lines suggest the absorber is likely a massive disk galaxy.

\par The other detection is an intervening system toward NVSS~J235421+455304, which is a quasar with a flux density of 1.8\,Jy at 1.4\,\unit{\giga\hertz} and located at a redshift of 1.992. This absorber was first reported by \citet{darlingDetection21Centimeter2004}, and re-detected by \citet{grashaEvolutionNeutralHydrogen2020}. \citet{darlingDetection21Centimeter2004} identified this system as a DLA based on the detection of the \mgii\ $\lambda\lambda 2796$ and $2803$, and Fe{\sc ii} $\lambda\lambda 2344$ and $2600$ absorption lines in the optical spectrum. As with NVSS~J031443+431405, no other NVSS sources of comparable brightness are present within the beam. However, the absorption feature is located at 798.2\,\unit{\mega\hertz}, only 1.8\,\unit{\mega\hertz} below CHIME's sampling rate at 800\,\unit{\mega\hertz}. Therefore, signals from frequencies above 800\,\unit{\mega\hertz} appear at lower frequencies in the measured data (aliasing), as the analog bandpass filter has a non-negligible response at frequencies slightly above 800\,\unit{\mega\hertz}. Since there is no absorption feature in the aliased contribution, this artificially increases the continuum level at 798.2\,\unit{\mega\hertz}. We correct for this contamination using the known bandpass filter response and the source's spectral index derived from cataloged flux measurements. After correction, we find a peak optical depth of $0.36\pm0.02$ from the fit.
\par
\begin{figure}
    \centering
    \includegraphics[width=1\linewidth]{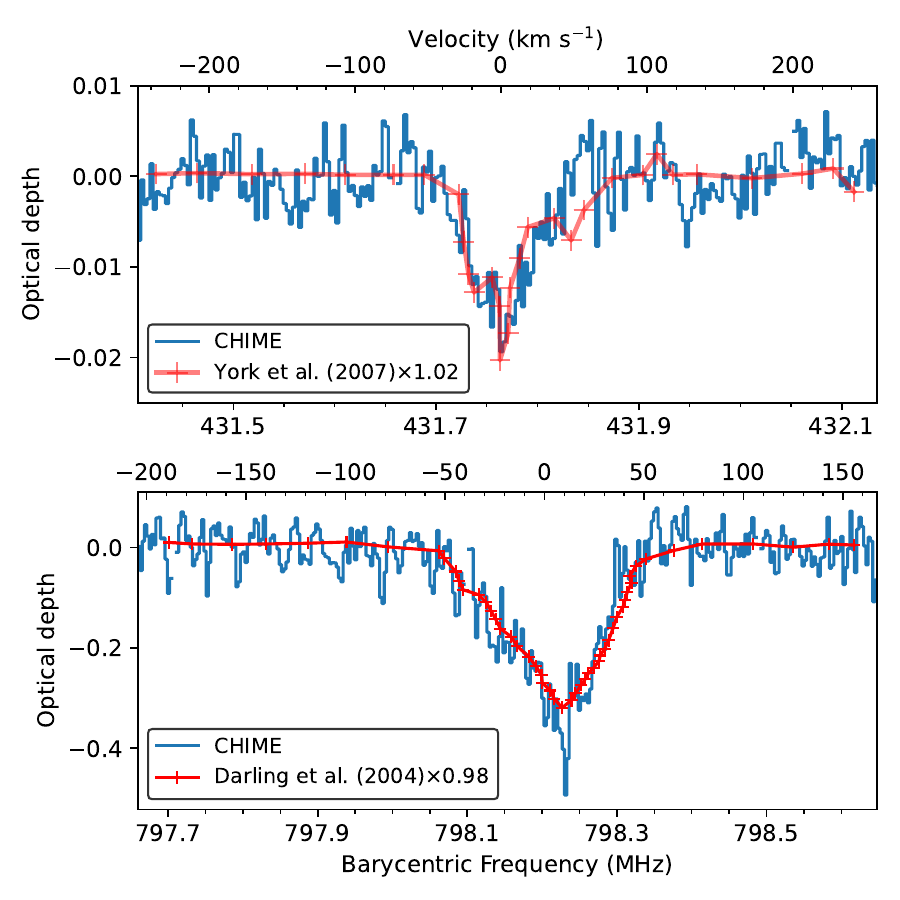}
    \caption{\textit{Top panel}: Absorption toward NVSS~J031443+431405 at z = 2.290 overplotted with digitized data from \cite{yorkDiscovery21cmAbsorption2007} (red). \textit{Bottom panel}: Absorption toward NVSS~J235421+455304 at z = 0.780 overplotted with digitized data from \cite{darlingDetection21Centimeter2004} (red).}
    \label{overplot}
\end{figure}

 
To quantify the agreement between our measurements and previous observations, we fitted the digitized data from \cite{yorkDiscovery21cmAbsorption2007} and \cite{darlingDetection21Centimeter2004} to our spectra using a scaling factor as the fitting parameter. Figure~\ref{overplot} shows the scaled previous observations plotted against our data. For NVSS~J031443+431405, we find that the CHIME spectrum matches the York \textit{et al.} data with a scaling factor of $1.02\pm0.08$.  For NVSS~J235421+455304, the best-fit scaling factor between our data and the Darling \textit{et al.} spectrum is $0.98\pm0.04$. Both comparisons show that CHIME successfully recovers the absorption profiles observed in previous studies, demonstrate excellent agreement between the datasets, and confirm the consistency of our measured optical depths.

\section{Summary and perspective}
We present the results of a pilot spectrally blind search for \tcm absorption systems along the lines of sight to 202 sources within a declination range of $35{\degree}$ to $60{\degree}$ using the CHIME telescope. The sources are at a redshift of $z > 0.77$, allowing them to serve as backlight for the absorbers. Using the CHIME Absorber pipeline, four datasets, each representing one month of data, are processed individually to produce spectra of 202 lines of sight. We then apply barycentric corrections, a search algorithm, and a validation algorithm as separate steps outside the pipeline to identify final candidates.  We detect two previously known intervening absorbers and one new associated system at a redshift of 2.32743. The new detection is only the fifth detection of an \tcm absorber associated with a quasar at redshift $z>2$. By modeling the absorption profile with a set of Gaussian functions, we estimate the optical depth, velocity-integrated optical depth, and $f \nhi / T_\mathrm{s}$ for our detections. 
\par The newly detected associated absorber toward NVSS~J164725+375218 exhibits a two-component Gaussian profile with a total velocity width of 69\,$\unit{\km.\s^{-1}}$ and a modest velocity offset of $\sim 300\, \unit{\km.\s^{-1}}$ relative to the host AGN redshift. These kinematic properties are consistent with \hi\ gas in a circumnuclear disk rather than outflowing material, adding to the small but growing sample of associated absorbers that can constrain models of the neutral gas environment in high-redshift AGNs.

\par The results presented here demonstrate CHIME's unique capabilities for \tcm absorption surveys. CHIME's continuous wide-area sky coverage at high spectral resolution, combined with its sensitivity to frequencies spanning 400 to 800 MHz, positions it as a powerful complement to existing surveys. While FLASH covers redshifts between 0.42 and 1.0 in the southern sky, MALS extends to $z<2$ at declinations $\delta<+40\degree$, and FASHI surveys $z<0.42$ at northern declinations, CHIME fills a critical gap at high redshifts ($0.78<z<2.55$) in the northern sky, a regime where detections of \tcm absorbers are scarce.
\par This pilot survey, while limited to sources with known optical redshifts, represents a genuinely spectrally blind search, because although we selected radio sources whose optical counterpart has a known redshift, we have not used any prior information about the expected \hi\ absorption frequency.   This distinguishes our approach from targeted surveys, where optical spectroscopy predetermines the search frequency range. However, requiring optical redshift information introduces a bias toward less dusty radio sources and those with strong UV emission. A fully blind survey must include all bright radio sources regardless of optical identification.

\par Looking forward, we plan to expand this work in several key directions. This includes processing additional months of archival data to improve sensitivity to fainter background sources, expanding the declination coverage beyond $35\degree$ to $60\degree$ to leverage CHIME's full field of view, and extending the survey to include radio sources without optical redshift information to eliminate the bias toward UV-bright AGNs and enable a truly dust-unbiased census of \tcm absorbers.  A comprehensive statistical analysis, including detailed calculations of the total absorption path probed by the survey, expected number of intervening and associated detections, and comparison with similar surveys at other redshifts (e.g., MALS and FLASH), will be presented alongside these expanded results in a forthcoming publication.

\begin{acknowledgements}
We thank the Dominion Radio Astrophysical Observatory, operated by the National Research Council Canada, for gracious hospitality and expertise. The DRAO is situated on the traditional, ancestral, and unceded territory of the syilx Okanagan people. We are fortunate to work on these lands.

\par CHIME is funded by grants from the Canada Foundation for Innovation (CFI) 2012 Leading Edge Fund (Project 31170), the CFI 2015 Innovation Fund (Project 33213), and by contributions from the provinces of British Columbia, Qu\'ebec, and Ontario. Long-term data storage and computational support for analysis is provided by Digital Research Alliance of Canada\footnote{\url{https://www.alliancecan.ca/en}} and SciNet\footnote{\url{https://www.scinethpc.ca/}}, and we thank their staff for flexibility and technical expertise that has been essential to this work, particularly Martin Siegert, Lixin Liu, and Lance Couture.

Additional support was provided by the University of British Columbia, McGill University, and the University of Toronto. CHIME also benefits from NSERC Discovery Grants to several researchers,  funding from the Canadian Institute for Advanced Research (CIFAR), and from the Dunlap Institute for Astronomy and Astrophysics at the University of Toronto, which is funded through an endowment established by the David Dunlap family.
\par Kiyoshi W. Masui holds the Adam J. Burgasser Chair in Astrophysics and received support from NSF grant 2008031. Ue-Li Pen is supported by the Natural Sciences and Engineering Research Council of Canada (NSERC), [funding reference number RGPIN-2019-06770, ALLRP 586559-23],  Canadian Institute for Advanced Research (CIFAR) and AMD AI Quantum Astro. Juan Mena-Parra acknowledges the support of an NSERC Discovery Grant (RGPIN-2023-05373). We acknowledge the support of the Natural Sciences and Engineering Research Council of Canada (NSERC), [funding reference number 569654].
\software{
bitshuffle \citep{2015bitshuffle},
caput \citep{caput},
ch\_pipeline \citep{ch_pipeline},
Cython \citep{Cython},
draco \citep{draco},
h5py \citep{h5py},
HDF5 \citep*{HDF5},
Matplotlib \citep{Matplotlib},
mpi4py \citep{mpi4py},
NumPy \citep{NumPy},
OpenMPI \citep{OpenMPI},
pandas (\citealt*{pandas}; \citealt{pandas_paper}),
peewee \citep{Peewee},
SciPy \citep{SciPy},
Skyfield \citep{Skyfield},
}

\end{acknowledgements}
\appendix
\twocolumngrid
\section{False positives and detections}
\label{sec:falsepositives}

Prior to visual inspection, the search algorithm in Section~\ref{sec:methods:search} yields 73 absorption candidates. Our visual inspection reduces this set down to the 3 detections we report in this work. In this section, we describe an alternative procedure that separates false positives from genuine detections. This method, which is based on a comparison between the distribution of positive and negative amplitude features, gives the same results as our visual inspection, supporting our usage of visual inspection as the primary method of discarding false positives. Such a method has previously been used in \tcm emission searches \citep{2012PASA...29..296S,2015MNRAS.448.1922S}, \tcm absorption searches \citep{2021MNRAS.503..985A}, and CO emission searches \citep{2016ApJ...833...67W,2018ApJ...864...49P,2020AJ....159..190L}.
\par Since emission-like features are not expected in our survey, positive amplitude features serve as a proxy for false positives in our analysis. To compare the distribution of positive and negative amplitude features, we re-run the Doppler-shift selection procedure from Section~\ref{sec:methods:search} separately for positive and negative amplitude features. For this test, we use peak-to-rms values rather than likelihood ratios to compare the distributions of positive and negative features. The peak-to-rms metric is more robust against band-shape residuals because it is normalized by the local spectral rms, whereas a likelihood ratio-based comparison would require a null model that explicitly incorporates the band-shape residuals. 
\par We calculate the rms for each resulting candidate by measuring the rms in four coarse channels surrounding the feature, excluding the coarse channel containing the candidate itself. Both the peak amplitude and the rms are measured from the spectrum smoothed by a rectangular boxcar whose width matches the best-fitting template from the matched filter.
\par
\begin{figure}
    \centering
    \includegraphics[width=1\linewidth]{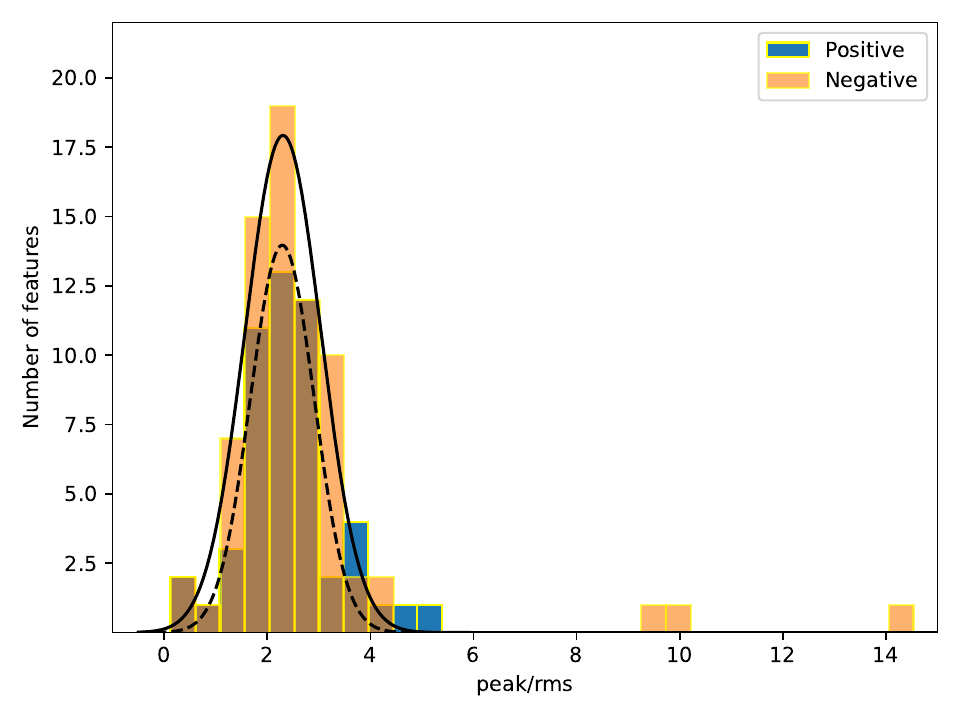}
    \caption{Number of features with positive and negative amplitudes as a function of their peak-to-rms value. Positive amplitude features represent emission-like signals, which are not expected in our survey and thus serve as a measure of false positives. Negative amplitude features follow the same trend, except for three outliers. Gaussian fits to the negative and positive feature distributions are shown by black solid and black dashed lines, respectively.  The Gaussian fit to the negative feature distribution confirms that the three absorption candidates that passed visual inspection are significant outliers to the distribution of false positives.}
    \label{false-positive}
\end{figure}

 Figure \ref{false-positive} presents the distribution of these features as a function of their peak-to-rms ratios. There are 51 positive and 73 negative features in our sample. We applied a two-sample Kolmogorov-Smirnov test comparing their distributions, which yields a KS statistic of $0.08$ and p-value of $0.98$, indicating the distributions are consistent with being drawn from the same population. 
We also separately fitted Gaussian functions to the distributions of positive and negative amplitude features. The fits yield similar mean and standard deviation for the peak-to-rms values for both distributions. This confirms that the majority of negative amplitude features are indeed false positives arising from similar effects that produce the positive features. Visual inspection of the 73 candidate spectra confirmed that all negative features, with the exception of three, are arising due to large residual band shape or due to  RFI contamination. The three candidates that successfully passed visual inspection criteria appear as clear outliers in Figure \ref{false-positive}, each exhibiting peak-to-rms ratios exceeding $10\sigma$, with $\sigma$ being the standard deviation of the fitted Gaussian.

\newpage
\bibliography{paper}{}
\bibliographystyle{aasjournalv7}

\end{document}